\documentclass[journal=jctc,manuscript=article,layout=traditional]{achemso}
\setkeys{acs}{articletitle = true}

\usepackage[version=3]{mhchem} 
\usepackage[T1]{fontenc}       

\usepackage{amsmath}
\usepackage{amsfonts}
\usepackage[theorems]{tcolorbox}
\usepackage{algorithm}
\usepackage{algpseudocode}
\usepackage{physics}
\usepackage{booktabs}
\usepackage{booktabs}
\usepackage{siunitx}
\usepackage[subrefformat=parens,labelformat=parens]{subcaption}
\usepackage{setspace}

\newcommand{\mrm}{\mathrm}
\newcommand{\mcl}{\mathcal}

\newcommand{\eps}[2]{\varepsilon^{#1}_{#2}}

\newcommand{\del}[2]{\delta_{#1,#2}}
\newcommand{\la}[2]{\lambda^{#1}_{#2}}
\newcommand{\bla}{\boldsymbol{\lambda}}
\newcommand{\La}{\Lambda}
\newcommand{\bLa}{\boldsymbol{\Lambda}}

\newcommand{\bOm}{\boldsymbol{\Omega}}
\newcommand{\btOm}{\boldsymbol{\tilde{\Omega}}}

\newcommand{\Ttpt}{T_{2+3}}
\newcommand{\bt}{\boldsymbol{\tau}}

\newcommand{\tBD}{\tilde{\boldsymbol{D}}}
\newcommand{\BD}{\boldsymbol{D}}
\newcommand{\tD}{\tilde{D}}
\newcommand{\bD}{\bar{D}}

\newcommand{\bR}{\boldsymbol{R}}
\newcommand{\bbR}{\bar{\boldsymbol{R}}}
\newcommand{\bL}{\boldsymbol{L}}
\newcommand{\bLT}{\boldsymbol{L}^T}
\newcommand{\bbL}{\bar{\boldsymbol{L}}}
\newcommand{\bzero}{\boldsymbol{0}}
\newcommand{\bI}{\boldsymbol{I}}
\newcommand{\bJ}{\boldsymbol{J}}
\newcommand{\bM}{\boldsymbol{M}}
\newcommand{\beT}{\boldsymbol{\eta}^T}

\newcommand{\half}{\frac{1}{2}}
\newcommand{\third}{\frac{1}{3}}

\newcommand{\hH}{\hat{H}}

\newcommand{\bbH}{\bar{\boldsymbol{H}}}

\newcommand{\tX}{\tilde{X}}

\newcommand{\Xnuo}{X_{\nu_{1}}}
\newcommand{\Xnut}{X_{\nu_{2}}}
\newcommand{\Xnur}{X_{\nu_{3}}}

\newcommand{\bs}[1]{\boldsymbol{#1}}

\newcommand{\Lcc}{\mathcal{L}_{\mathrm{CC}}}

\newcommand{\Lccsdt}{\mathcal{L}_{\mathrm{CCSDT}}}
\newcommand{\Lcct}{\mathcal{L}_{\mathrm{CC3}}}

\newcommand{\rbra}[1]{\bra{\mathrm{#1}}}
\newcommand{\rbrat}[1]{\langle\tilde{\mathrm{#1}}|}
\newcommand{\rket}[1]{\ket{\mathrm{#1}}}
\newcommand{\rbraket}[2]{\braket{\mathrm{#1}}{\mathrm{#2}}}

\newcommand{\brat}[1]{\bra{#1}}
\newcommand{\lref}[0]{\rbra{\phi_{0}}}
\newcommand{\rref}[0]{\rket{\phi_{0}}}

\newcommand{\peq}{\mathrel{+}=}
\newcommand{\meq}{\mathrel{-}=}
\newcommand{\ssum}[2]{\sum_{\substack{#1 \\ #2}}}

\newcommand{\dder}[3]{\frac{\partial^2{#1}}{\partial{#2}\partial{#3}}}

\newcommand{\ta}[2]{\tau^{#1}_{#2}}
\newcommand{\tta}[2]{\tilde{\tau}^{#1}_{#2}}

\newcommand{\Lx}[2]{L^{#1}_{#2}}
\newcommand{\tLx}[2]{\tilde{L}^{#1}_{#2}}

\newcommand{\Om}[2]{\Omega^{#1}_{#2}}
\newcommand{\tOm}[2]{\tilde{\Omega}^{#1}_{#2}}
\newcommand{\tZo}[1]{\tilde{Z}^\text{o}_{#1}}
\newcommand{\tZv}[1]{\tilde{Z}^\text{v}_{#1}}

\newcommand{\nv}[1]{n^{#1}_{\mathrm{V}}}
\newcommand{\no}[1]{n^{#1}_{\mathrm{O}}}

\newcommand{\mc}[3]{\multicolumn{#1}{#2}{#3}}
\newcommand{\nan}[0]{\mc{1}{c}{-}}

\newcommand{\dalton}{\textsc{Dalton}}
\newcommand{\psif}{\textsc{Psi4}}
\newcommand{\cfour}{\textsc{Cfour}}

\newcommand{\ncall}{n_{\mrm{calls}}}

\newtcolorbox{mymathbox}[2][]{ams align, colback=white, sharp corners,
                              title=#2, #1}

\makeatletter
\def\lefteqno{\tagsleft@true}\def\righteqno{\tagsleft@false}
\makeatother

\author{Alexander C. Paul}
\altaffiliation{Contributed equally to this work}
\author{Rolf H. Myhre}
\altaffiliation{Contributed equally to this work}
\author{Henrik Koch}
\email{henrik.koch@sns.it}
\affiliation{Department of Chemistry, Norwegian University of Science and Technology,
             NTNU, 7491 Trondheim, Norway}
\alsoaffiliation{Scuola Normale Superiore, Piazza dei Cavaleri 7, 56126 Pisa, Italy}

\title[CC3]
{A new and efficient implementation of CC3}

\abbreviations{CC, CC3}

\begin{document}

\begin{tocentry}
   \includegraphics[scale=0.42]{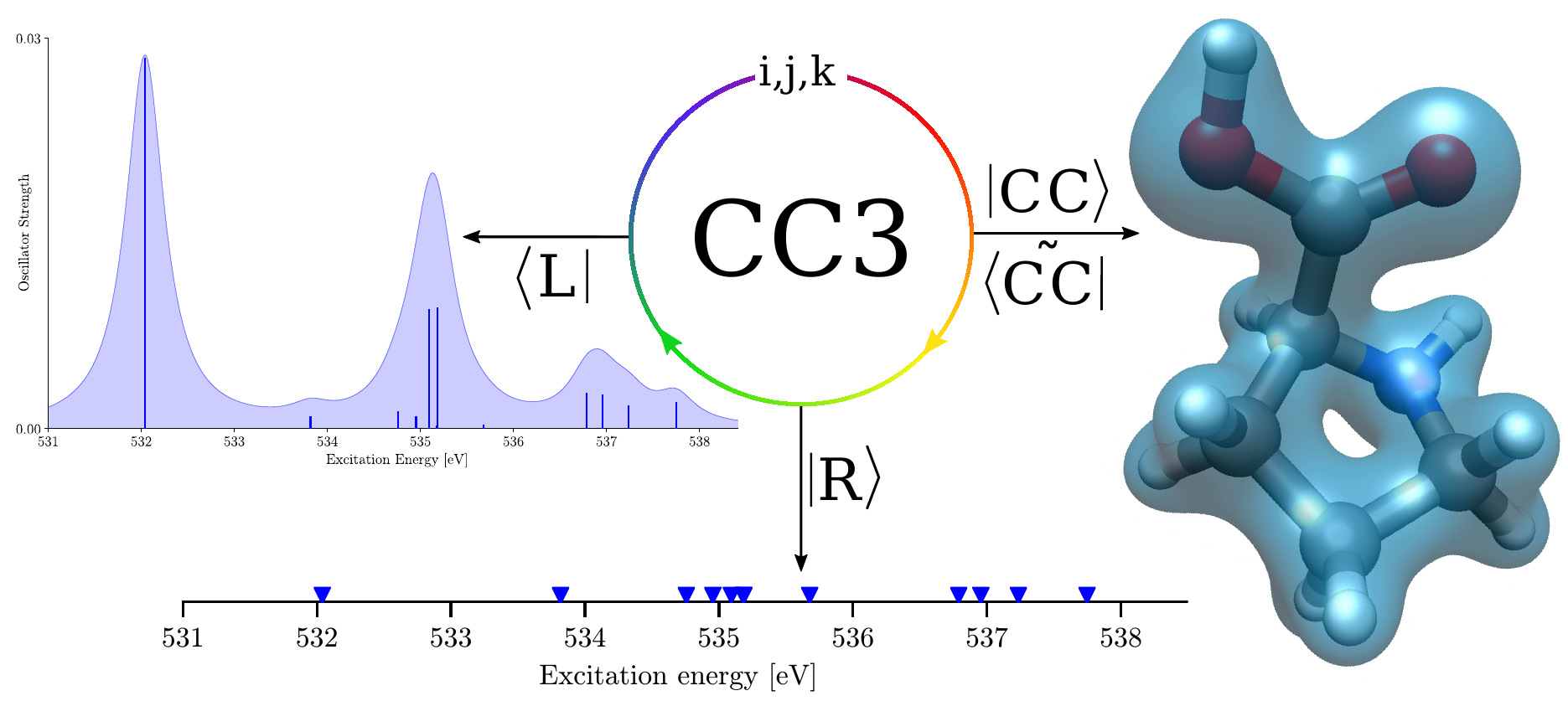}
\end{tocentry}


\begin{abstract}
   We present a new and efficient implementation of the closed shell coupled cluster singles
   and doubles with perturbative triples method (CC3) in the electronic structure program $e^T$.

   Asymptotically, 
   a ground state calculation has an iterative cost of $4\nv{4}\no{3}$ 
   floating point operations (FLOP),
   where $\nv{}$ and $\no{}$ are the number of virtual and occupied orbitals respectively.
   The Jacobian and transpose Jacobian transformations,
   required to iteratively solve for excitation energies and transition moments,
   both require $8\nv{4}\no{3}$ FLOP.
   We have also implemented equation of motion (EOM) transition moments for CC3.
   The EOM transition densities require recalculation of triples amplitudes,
   as $\nv{3}\no{3}$ tensors are not stored in memory.
   This results in a noniterative computational cost 
   of $10\nv{4}\no{3}$ FLOP for the ground state density
   and $26\nv{4}\no{3}$ FLOP per state for the transition densities.
   The code is compared to the CC3 implementations in \cfour{},
   \dalton{} and \psif{}.
   We demonstrate the capabilities of our implementation
   by calculating valence and core excited states of L-proline.
\end{abstract}

\section{Introduction} \label{intro}

X-ray spectroscopies such as near edge X-ray absorption fine structure (NEXAFS)
can provide a detailed insight into the electronic structure of molecules
and their local environment\cite{xray1, Scholz}.
With the new facilities at the European XFEL and LCLS2 at SLAC 
the number of high resolution spectroscopic experiments is increasing. 
Accurate modelling is a great aid when interpreting the spectroscopic data,
providing new insights into the underlying chemistry.
However, 
modelling the high energy excitations measured in X-ray spectroscopy is challenging 
because they typically generate a core hole 
which in turn results in a large contraction of the electron density. 
To accurately describe this contraction, 
one either has to include triple excitations 
or explicit excited state orbital relaxation 
in the wavefunction description\cite{Norman:ChemRev:18,nitro, LanBenchmark,NOCIcore}.

Coupled cluster theory is the preferred model for calculating
spectroscopic properties for molecules,
combining high accuracy and correct scaling with system size
in the coupled cluster response theory (CCRT)
formulation\cite{WavFuncRev,CCRespFunc, CCRespRev}.
Coupled cluster singles and doubles (CCSD) is the most widely used
variant of coupled cluster,
due to its high accuracy
and relatively feasible computational scaling as 
$\mcl{O}(\nv{4}\no{2})$,
where $\nv{}$ is the number of virtual and $\no{}$ is the number of occupied orbitals.
Nevertheless,
for some properties like core excitation energies,
CCSD can deviate by several electron volts from experimental values.
These deviations are reduced by an order of magnitude if triples 
are included in the description of the wave function \cite{glycine, nitro}.
However,
coupled cluster singles, doubles and triples (CCSDT) is
usually unfeasible due to the $\nv{3}\no{3}$ memory requirement
and $\mcl{O}(\nv{5}\no{3})$ computational cost.
Approximating the triples amplitudes can reduce the computational cost to $4\nv{4}\no{3}$
floating point operations (FLOP) and the required memory to $\nv{2}\no{2}$.
Note that this is twice the scaling usually reported in the literature
because a matrix-matrix multiplication involves an addition and a multiplication.

Approximate triples models are typically categorized as noniterative 
and iterative models.
For the noniterative models,
a triples energy correction is computed after solving the CCSD equations.
The terms included in the energy correction are usually determined
based on a many-body perturbation theory (MBPT) like expansion 
of the energy\cite{MollerPlesset, MBPT, CCSDpt4, CCSDbt}.
However, 
CCSD(T),
by far the most popular of these methods, 
does not follow a strict MBPT expansion of the energy.
For CCSD(T) the energy is expanded consistently to fourth order
and one additional fifth order term is added\cite{CCSD(T)}.
It was later shown that CCSD(T) can be viewed as an MBPT like expansion 
from the CCSD wave function\cite{rationalisation}.
Similar approaches have also been proposed for excitation energies\cite{exciptsd,ccsdptpa}.
A related method is the $\Lambda$CCSD(T) method 
where the parameters of the left wave function are included 
in the MBPT expansion\cite{LambdaCCSDpT2, LambdaCCSDpT}.
The completely renormalized CCSD(T) method, 
intended for multireference states has also been extended to excited states\cite{CREOMptsd}.
Other models include CCSDR3\cite{CCSDR3} and EOM-CCSD$^*$ \cite{ccstar1, ccstar2}, 
where a triples correction is added to the CCSD excitation energies. 
The iterative methods are generally more computationally expensive 
than the noniterative methods,
but they are usually more accurate.
The CCSDT-n models\cite{CCSDTn}
and CC3\cite{CC3,mlcc3} are the most well known of these methods. 
The two models have the same computational cost, 
but CC3 is more accurate due to the full inclusion of single excitation amplitudes.\cite{CCexcienergies}
Recently, 
CC3 ground and excited states were combined with the pair natural orbital approximation 
in order to extend the model to larger systems\cite{PNOCC3}.
For a more extensive discussion of approximate triples methods and their accuracy,
see Ref.\,\citenum{CChighexci, CCaccuracy1, CCaccuracy2, CCaccuracy3, CCaccuracy4, CCaccuracy5}.

Due to the high computational cost,
an efficient CC3 implementation is required for larger molecules\cite{CC3, mlcc3}.
In this paper we present an implementation of CC3 ground and excited states,
as well as equation of motion (EOM)\cite{EOMCC} transition moments.
Although the EOM formalism has been shown to be less accurate than CCRT 
for transition moments\cite{CCRespFunc},
the differences are believed to be small for high level methods like CC3.\cite{size_int_trans, EOMsize}
The current implementation has an iterative cost of $4\nv{4}\no{3}$ FLOP for the ground state
and $8\nv{4}\no{3}$ FLOP for excited states.
For comparison, 
the old CC3 excited states implementation in \dalton{} has an iterative cost of $30\nv{4}\no{3}$ FLOP
and the new implementation in \dalton{} requires $10\nv{4}\no{3}$ FLOP per iteration.\cite{mlcc3}
Note that it is erroneously stated in the literature
that the minimal computational cost is $10\nv{4}\no{3}$ FLOP per iteration.\cite{CC3imp}
For core excited states,
we use the core valence separation (CVS) approximation\cite{CVS, CVS_CC,CVS_CC_erratum}.
This reduces the iterative computational cost to $8\nv{4}\no{2}$ FLOP for excitation energies,
however,
the computational cost of the ground state calculation remains unchanged.

\section{Theory} \label{theory}

In this section,
we will derive the equations for closed shell CC3 within the EOM formalism.
Note that almost all the equations in this section 
are equally applicable in the open shell case by changing the definitions 
of the Hamiltonian and the one-electron operator. 
Consider the coupled cluster wave function,

\begin{equation}\label{ccwave}
   \rket{CC} = e^{T}\rref \quad T = \sum_{\mu} \tau_\mu X_\mu.
\end{equation}

Here, $\rref$ is a canonical reference Slater determinant,
usually the Hartree-Fock wave function,
and $T$ is the cluster operator with $\mu$ labeling unique excited determinants.
The excitation operator, 
$X_\mu$, 
maps the reference,
$\rref$
into determinant $\rket{\mu}$ and
$\tau_\mu$ is the corresponding parameter,
referred to as an amplitude.
In the closed shell case, $X_\mu$ is defined as a string of 
the standard singlet excitation operators, 
$E_{ai}$, 
with a corresponding normalization factor. 
For example, 
a double excitation operator is given by 
\begin{equation}\label{doubexci}
   X^{ab}_{ij} = E_{ai}E_{bj},
\end{equation}
and we use the standard notation,
where the indices $i,j,k\dots$ refer to occupied,
$a,b,c\dots$ to virtual and $p,q,r\dots$ to general orbitals.
We will work in a biorthonormal basis and define a contravariant excitation operator, 
$\tX_\mu$,
so that the left space is spanned by determinants biorthonormal to the right.\cite{Bibel}
\begin{equation}\label{cocontra}
   \rbra{\mu} = \lref\tX_\mu \quad 
   \rket{\nu} = X_\nu\rref \quad 
   \rbraket{\mu}{\nu} = \del{\mu}{\nu}
\end{equation} 

In order to obtain the ground state energy,
we introduce a biorthonormal parametrization for the left state,
\begin{equation}\label{ccleft}
   \rbrat{CC} = \lref(1 + \La)e^{-T} \quad \La = \sum_{\mu\ne\phi_0} \lambda_\mu \tX_\mu.
\end{equation}
Inserting these expressions
into the Schr\"odinger equation,
we obtain the coupled cluster Lagrangian,
\begin{equation}\label{ccL}
      \Lcc = \rbrat{CC}\hH\rket{CC}
           = \lref(1 + \La)e^{-T}\hH e^{T}\rref,
\end{equation}
where $\hH$ is the electronic Hamiltonian\cite{Bibel},
\begin{equation}
    \hH = \sum_{pq} h_{pq} E_{pq} 
        + \half \sum_{pqrs} g_{pqrs} (E_{pq}E_{rs} - E_{ps} \delta_{qr}).
\end{equation}
The equations for full configuration interaction (FCI) are recovered from this Lagrangian
if the excitation space is not truncated.
The biorthonormal left side is then equivalent to the conjugate
of the right side up to a normalization factor.

Determining the stationary points of $\Lcc$ results in 
the equations for the parameters $\bt$ and $\bla$.
The derivatives with respect to $\bla$ 
give the familiar coupled cluster projection equations for the
amplitudes and the derivatives with respect to $\bt$ give the equations for $\bla$.
In practice,
$T$ and $\La$ are truncated at some excitation level with respect to the
reference determinant.
For example, the CCSDT cluster operators are defined as the sum of the 
singles, 
doubles, 
and triples cluster operators. 

\begin{equation}\label{T3}
   T^{CCSDT} = T_1 + T_2 + T_3 \quad \La^{CCSDT} = \La_1 + \La_2 + \La_3
\end{equation}

The $\exp(T_1)$ operator can be viewed as a biorthogonal orbital transformation and
we employ the $T_1$-transformed Hamiltonian throughout,
\begin{equation}\label{Ht1}
    H = e^{-T_1}\hH e^{T_1}.
\end{equation}
Note that we do not use the standard notation to avoid over dressing of the operators. 
The equations for CCSDT then become those of CCDT.
Inserting these definitions into $\Lcc$,
we get the CCSDT Lagrangian,

\begin{equation}\label{Lccsdt}
   \begin{aligned}
      \Lccsdt &= \lref H + [H,T_2]\rref\\
              &+ \sum_{\mu_1}\la{}{\mu_1}\brat{\mu_1}H
                                                   + [H,T_2]
                                                   + [H,T_3]\rref\\
              &+ \sum_{\mu_2}\la{}{\mu_2}\brat{\mu_2}H
                                                   + [H,T_2]
                                                   + \half[[H,T_2],T_2]
                                                   + [H,T_3]\rref\\
              &+ \sum_{\mu_3}\la{}{\mu_3}\brat{\mu_3}[H,T_2]
                                                   + \half[[H,T_2],T_2]
                                                   + [H,T_3]
                                                   + [[H,T_2],T_3]\rref.\\
   \end{aligned}
\end{equation}

The last two commutator terms of eq\,\eqref{Lccsdt} make the cost of 
the full CCSDT model scale as $\mcl{O}(\nv{5}\no{3})$.
To reduce the cost we use a perturbation scheme\cite{CC3,mlcc3},
where the transformed Hamiltonian
is divided into an effective one particle operator
and a fluctuation potential,
similar to MBPT,\cite{MollerPlesset, MBPT}
\begin{equation}\label{Hsplit}
   H = F + U.
\end{equation}
The operators are assigned orders as summarized in Table\,\ref{orders}.
\begin{equation}\label{Lcc3}
   \begin{aligned}
      \Lcct &= \sum_{\mu_1}\la{}{\mu_1}\brat{\mu_1}H
                                                 + [H,T_2]
                                                 + [H,T_3]\rref\\
            &+ \sum_{\mu_2}\la{}{\mu_2}\brat{\mu_2}H
                                                 + [H,T_2]
                                                 + \half[[H,T_2],T_2]
                                                 + [H,T_3]\rref\\
            &+ \sum_{\mu_3}\la{}{\mu_3}\brat{\mu_3}[H,T_2]
                                                 + [F,T_3]\rref\\
   \end{aligned}
\end{equation}
The CC3 Lagrangian, 
eq\,\eqref{Lcc3},
is obtained by discarding terms from the CCSDT Lagrangian,
that are of fifth order in the perturbation and higher, 
assuming a canonical basis.
The singles amplitudes, 
both in $\La_1$ and $T_1$, 
are considered to be zeroth order in the perturbation 
as they are viewed as approximate orbital transformation parameters\cite{NOCC,OACC}.
In contrast to MBPT where the first contribution 
of the single excitations appears in second order.

\begin{table}
   \centering
   \caption{Perturbation orders for CC3.
            $F_{\text{oo}}$ and $F_{\text{vv}}$ refer to the diagonal blocks of the Fock matrix, 
            while $F_{\text{vo}}$ and $F_{\text{ov}}$ refer to the off diagonal blocks.
            $T$ and $\Lambda$ refer to ground state parameters.
            $r$, $l$, $L$ and $R$ refer to EOM parameters\label{orders}.}
   \begin{tabular}{cccc}
      \toprule
      Order        & 0                      & 1              & 2\\
      \midrule
      Hamiltonian  & $F_{\text{oo}}$, $F_{\text{vv}}$ & $F_{\text{vo}}$, $F_{\text{ov}}$, $U$\\
      Ground state & $\La_{\mu_1}$, $T_{\mu_1}$ 
                   & $\La_{\mu_2}$, $T_{\mu_2}$ 
                   & $\La_{\mu_3}$, $T_{\mu_3}$ \\
      EOM          & $r$, $l$, $L_{\mu_1}$, $R_{\mu_1}$ 
                   & $L_{\mu_2}$, $R_{\mu_2}$   
                   & $L_{\mu_3}$, $R_{\mu_3}$\\
      \bottomrule
   \end{tabular}
\end{table}

In coupled cluster theory, 
excitation energies and other spectroscopic properties
are usually computed using either CCRT or the EOM formalism.
In CCRT, 
time dependent expectation values of molecular properties are expanded in orders 
of a frequency dependent perturbation.
The frequency dependent expansion terms are referred to as response functions and 
excitation energies and transition moments 
are determined from the poles and residues of the linear response function.
In EOM theory,
the starting point is a CI parametrization for the excited states.
The eigenvalue problem for the Hamiltonian in this basis
gives excited states and excitation energies.
Coupled cluster response theory and EOM give the same expressions 
for the excitation energies, 
however,
the transition moments differ.

To solve the EOM equations, 
the similarity transformed Hamiltonian is projected onto the reference 
and the truncated excitation space resulting in the Hamiltonian matrix,
\begin{equation}\label{bbw}
   \bbH_{\mu\nu} = \brat{\mu}e^{-T}\hH e^{T}\ket{\nu}.
\end{equation}
This matrix is not symmetric,
hence, the left and right eigenvectors will not be Hermitian conjugates,
but they will be biorthonormal.
\begin{equation}
   \bbH\bR_m = E_m\bR_m
   \quad
   \bL^T_m\bbH = E_m\bL^T_m
   \quad
   \bL^T_m\bR_n = \del{m}{n}
\end{equation}
We introduce the convenient notation
\begin{equation}
   \bR_m =
   \begin{pmatrix}
      r_m\\
      \bbR_m
   \end{pmatrix}
   \quad
   \bL_m =
   \begin{pmatrix}
      l_m\\
      \bbL_m
   \end{pmatrix},
\end{equation}
where $l_m$ and $r_m$ refer to the first element of the vectors
and $\bbL_m$ and $\bbR_m$ refer to the rest.
The vectors $\bL_m$ and $\bR_m$ correspond to the operators $L_m$ and $R_m$,
which have a similar form as $\La$ and $T$,
but also include reference contributions.

The EOM excited right states are written as
\begin{equation}\label{EOMright}
   \rket{m} = R_m\rket{CC} = e^{T} R_m\rref,
\end{equation}
and the left states are written as
\begin{equation}\label{EOMleft}
   \rbra{m} = \lref L_m e^{-T}.
\end{equation}
Because the $\bt$ amplitudes are solutions to the coupled cluster ground state equations,
the first column of $\bbH$ is zero,
except for the first element which equals the ground state energy, $E_0$, 
and the eigenvalues of $\bbH$ correspond to the energies of the EOM states.
\begin{equation}\label{bbHform}
   \bbH =
   \begin{pmatrix}
      E_0    & \beT\\
      \bzero & \bM
   \end{pmatrix}
\end{equation}

In the following,
the index $m$ will refer to states other than the ground state, 
which is denoted by $0$.
From the structure of the Hamiltonian matrix,
we see that the vector $\bR_0$,
with the elements $r_0 = 1$ and $\bbR_0 = \bzero$,
corresponds to the ground state.
For the right excited states,
$\bbR_m$ must be an eigenvector of $\bM$ with eigenvalue $E_m$.
Similarly,
for the left excited states $l_m = 0$ and $\bbL_m$ has to be a left eigenvector of $\bM$,
due to the biorthonormality with $\bR_0$ and $\bR_m$.
The left ground state, 
$\bL_0$, 
has the component $l_0 = 1$ and the vector $\bbL_0$ 
is obtained from the linear equation
\begin{equation}
   \beT = \bbL^T_0(E_0\bI - \bM),
\end{equation}
where $\bI$ is the identity matrix.
Finally, 
$r_m = -\bbL^T_0\bbR_m$
to ensure biorthogonality between $\bR_m$ and $\bL_0$.
The matrix $\bJ = (\bM - E_0\bI)$ is the derivative of the Lagrangian
with respect to $\bt$ and $\bla$,
\begin{equation}
   \bJ_{\mu\nu} = \dder{\Lcc}{\lambda_\mu}{\tau_\nu},
\end{equation}
and is called the Jacobian.
As required, 
the equation for $\bL_0$ is the same as for $\La$.
The CCSDT Jacobian is given by
\footnotesize
\begin{equation}\label{jmatrix}
\begin{aligned}
   &\bJ_{CCSDT} =
   \\
   &
   \begin{pmatrix}
      \rbra{\mu_1}[H + [H,T_2],\Xnuo]\rref &
      \rbra{\mu_1}[H,\Xnut]\rref &
      \rbra{\mu_1}[H,\Xnur]\rref
      \\
      \rbra{\mu_2}[H + [H, \Ttpt],\Xnuo]\rref &
      \rbra{\mu_2}[H + [H,T_2],\Xnut]\rref &
      \rbra{\mu_2}[H,\Xnur]\rref
      \\
      \rbra{\mu_3}[H + [H,\Ttpt] + \half[[H,T_2],T_2],\Xnuo]\rref &
      \rbra{\mu_3}[H + [H,\Ttpt],\Xnut]\rref &
      \rbra{\mu_3}[H + [H,T_2],\Xnur]\rref
   \end{pmatrix}
\end{aligned}
\end{equation}
\normalsize
and the CCSDT $\bs{\eta}$ vector is given as
\begin{equation}\label{bvector}
   \beT_{CCSDT} =
   \begin{pmatrix}
      \lref[H,\Xnuo]\rref &
      \lref[H,\Xnut]\rref &
      \bzero^T
   \end{pmatrix},
\end{equation}
where $\Ttpt$ is shorthand notation for $T_2 + T_3$.
These expressions are written in commutator form
which requires that the projection equations for $T$ are satisfied.

For EOM CC3 we introduce a perturbation expansion.
Our starting point is the expression for the energy of the EOM states,
\begin{equation}
   E_m = \bLT_m\bbH\bR_m.
\end{equation}
We assign the same perturbation orders to $\bL$ and $\bR$ as to $T$ and $\La$,
see Table\,\ref{orders}.
As CC3 does not satisfy the projection equations,
the first column of $\bbH$ will not be zero after the first element.
However, discarding the terms that are fifth order or higher, 
we are left with the expressions for the CC3 ground state residuals which are zero.
In order to derive the correct CC3 Jacobian,
known from CCRT\cite{CC3resp},
we discard terms from the CCSDT Jacobian in commutator form
using perturbation theory, giving
\footnotesize
\begin{equation}\label{jcc3}
\begin{aligned}
   &\bJ_{CC3} =
   \\
   &
   \begin{pmatrix}
      \rbra{\mu_1}[H + [H,T_2],\Xnuo]\rref &
      \rbra{\mu_1}[H,\Xnut]\rref &
      \rbra{\mu_1}[H,\Xnur]\rref
      \\
      \rbra{\mu_2}[H + [H, \Ttpt],\Xnuo]\rref &
      \rbra{\mu_2}[H + [H,T_2],\Xnut]\rref &
      \rbra{\mu_2}[H,\Xnur]\rref
      \\
      \rbra{\mu_3}[H + [H,T_2],\Xnuo]\rref &
      \rbra{\mu_3}[H,\Xnut]\rref &
      \rbra{\mu_3}[F,\Xnur]\rref
   \end{pmatrix}.
\end{aligned}
\end{equation}
\normalsize

To obtain EOM biorthogonal expectation values, 
the biorthogonal states are inserted 
into the expressions for the CI expectation values.
For a given one-electron operator, $A=\sum_{pq}A_{pq}E_{pq}$,
the biorthogonal expectation values are expressed 
in terms of left and right transition density matrices,\cite{EOMdens1, EOMdens2}
$\tBD^{n,m}$ and $\BD^{n,m}$:
\begin{equation}
    \rbrat{CC}A\rket{m}\rbra{m}A\rket{CC} = 
    \Big(\sum_{pq}\tD^{0,m}_{pq}A_{pq}\Big)
    \Big(\sum_{pq}D^{m,0}_{pq}A_{pq}\Big).
\end{equation}
The elements of the right transition density are defined as:
\begin{equation}\label{D}
   D^{m,0}_{pq} = \rbra{m}E_{pq}\rket{CC},
\end{equation}
while the elements of $\tBD^{0,m}$ are given by
\begin{equation}\label{tD}
   \begin{aligned}
      \tD^{0,m}_{pq} &= \rbrat{CC}E_{pq}\rket{m}\\
                     &= \lref(1+\La)e^{-T}E_{pq}e^{T}R_m\rref\\
                     &= \lref(1+\La)e^{-T}E_{pq}e^{T}\bar{R}_m\rref\\
                     &+ r_m\lref(1+\La)e^{-T}E_{pq}e^{T}\rref\\
                     &= \bD^{0,m}_{pq} + r_m D^{0,0}_{pq},
   \end{aligned}
\end{equation}
where $\BD^{0,0}$ is the ground state density.

\section{Implementation}\label{implementation}

The closed shell CC3 ground state,
singlet excitation energies,
and EOM transition moments have been implemented
in the $e^T$ program.\cite{eT}
The core part of the algorithms is a triple loop over the occupied indices
$i \geq j \geq k$ as proposed for CCSD(T) by Rendell
\textit{et al.}\cite{fasttriples}
and has been used in several other implementations.\cite{TandQ, mlcc3, spinQuad}.
Within the triple loop we first construct the triples amplitudes
for a given set of $ \{i,j,k\} $ and contract them with
integrals to obtain the contribution to the resulting vector.
By restricting the loop indices
and exploiting the permutational symmetry,
\begin{equation}
     \tau^{abc}_{ijk} = \tau^{bac}_{jik} = \tau^{cba}_{kji}
   = \tau^{acb}_{ikj} = \tau^{cab}_{kij} = \tau^{bca}_{jki},
\end{equation}
the computational cost of constructing the triples amplitudes
is reduced by a factor of six.
An outline of the algorithm to construct the triples contribution 
to the ground state residual, 
$\bOm$, 
is given in Algorithm\,\ref{alg_omega}.
Integrals in $T_1$-transformed basis are denoted by $g_{pqrs}$.
The equation for the triples amplitudes
includes a permutation operator, 
defined by
\begin{equation}\label{permutation}
    P^{abc}_{ijk}B^{abc}_{ijk} 
    = B^{abc}_{ijk} + B^{bac}_{jik} + B^{cba}_{kji}
    + B^{acb}_{ikj} + B^{bca}_{jki} + B^{cab}_{kij}
\end{equation}
and the orbital energy difference
\begin{equation}
   \eps{abc}{ijk} = \eps{}{a}+\eps{}{b}+\eps{}{c}-\eps{}{i}-\eps{}{j}-\eps{}{k},
\end{equation}
where $\eps{}{p}$ is the energy of orbital $p$.
To recover all contributions to the $\bOm$ vector from the restricted loops, 
all unique permutations of $i,j,k$ have to be considered.
This results in six terms when all the occupied indices are unique 
and three terms in the case of two occupied indices being equal.
If all three occupied indices are identical, 
there is no contribution,
as this corresponds to a triple excitation from a single orbital.
In order to avoid reading two-electron integrals from file inside the loop,
the program checks if all the integrals can be kept in memory,
otherwise they are read in batches of $i,j,k$ in additional outer loops.
To minimize reordering inside the loop and ensure efficient matrix contractions,
the integrals are reordered and written to disk before entering the loop.

Asymptotically, 
reordering of the amplitudes or making linear combinations of them scale as $\nv{3}\no{3}$. 
However, 
these operations are typically memory bound. 
For example, 
reordering the amplitudes from 123 to 312 ordering 
took 57 seconds 
while the fastest $\nv{4}\no{3}$ matrix multiplication took 240 seconds 
for a system with 431 virtual and 29 occupied orbitals. 
The calculation was run on a node with two Intel Xeon-Gold 6138 2.0 GHz CPUs with 20 cores each
and 320 GB of memory. 
Reordering times are highly dependent on hardware and compiler, 
but it is clear that they are significant and
constructing linear combinations is even more time consuming. 
By constructing contravariant triples amplitudes, 
\begin{equation}\label{contra_t}
   \tta{abc}{ijk} = 4\ta{abc}{ijk} - 2\ta{bac}{jik} - 2\ta{cba}{kji}
                  - 2\ta{acb}{ikj} + \ta{cab}{kij} + \ta{bca}{jki},
\end{equation}
no additional linear combinations are required to construct 
the contravariant residual $\btOm$.
This residual can then be transformed back to the covariant residual outside the loop. 
\begin{equation}
   \tOm{a}{i} = \Om{a}{i}
\end{equation}
\begin{equation}
   \tOm{ab}{ij} = 2\Om{ab}{ij} - \Om{ba}{ij}, \quad \Om{ab}{ij} = \frac{1}{3}(2\tOm{ab}{ij} + \tOm{ba}{ij}),
\end{equation}

For systems with spatial symmetry, 
considerable savings could be achieved by taking symmetry into account, 
both in computational cost and memory. 
However, 
this results in greatly increased complexity of the code
and spatial symmetry is most relevant for small molecular systems. 
Consequently, 
it is not exploited in our implementation. 

\begin{algorithm}
\caption{Algorithm to construct the CC3 ground state equations.}
\label{alg_omega}
\begin{algorithmic}\onehalfspacing
    \While{not converged}
      \For{i = 1, $\no{}$}
         \For{j = 1, i}
            \For{k = 1, j}

               \State $\ta{abc}{ijk} \gets -(\eps{abc}{ijk})^{-1} P^{abc}_{ijk}
                          \Big(\sum_d \ta{ad}{ij}g_{bdck} -
                               \sum_l \ta{ab}{il}g_{ljck}\Big)$

               \State $\tta{abc}{ijk} \gets 4\ta{abc}{ijk}
                                          - 2\ta{acb}{ijk}
                                          - 2\ta{cba}{ijk}
                                          - 2\ta{bac}{ijk}
                                          +  \ta{bca}{ijk}
                                          +  \ta{cab}{ijk}$

               \For{Permutations of $i,j,k$}
                  \State $\tOm{a\phantom{b}}{i} \peq \sum_{bc}\tta{abc}{ijk}g_{jbkc}$
                  \State $\tOm{ab}{ij} \peq \sum_{c}\tta{abc}{ijk}F_{kc}$
                  \State $\tOm{ab}{il} \meq \sum_{c}\tta{abc}{ijk}g_{jlkc}$
                  \State $\tOm{ad}{ij} \peq \sum_{bc}\tta{abc}{ijk}g_{dbkc}$
               \EndFor

            \EndFor
         \EndFor
      \EndFor
      \State $\Om{ab}{ij} \peq \third P^{ab}_{ij} (2\tOm{ab}{ij} + \tOm{ba}{ij})$
    \EndWhile
\end{algorithmic}
\end{algorithm}

For excited state calculations, 
we may reduce the iterative cost
from $10\nv{4}\no{3}$ to $8\nv{4}\no{3}$ FLOP by constructing 
$\bt_{3}$-dependent intermediates before entering the iterative loop.
This is carried out in a preparation routine 
outlined in Algorithm\,\ref{alg_prep}.
The same intermediates are used in the algorithms for both $\bL$ and $\bR$.
Nevertheless, 
we still have to construct the $\bt_{3}$ amplitudes in each iteration,
see supporting information.
In theory 
it would be possible to construct an intermediate of size $\nv{3}\no{3}$ for this term as well, 
reducing the iterative computational cost to $6\nv{4}\no{3}$ FLOP.
However,
this intermediate would cost $2\nv{4}\no{4}$ FLOP to construct.

\begin{algorithm}
\caption{Preparation for the CC3 Jacobian transformations.}
\label{alg_prep}
\begin{algorithmic}\onehalfspacing
   \For{i = 1, $\no{}$}
      \For{j = 1, i}
         \For{k = 1, j}

            \State $\ta{abc}{ijk} \gets - (\eps{abc}{ijk})^{-1} P^{abc}_{ijk}
                   \Big(\sum_d \ta{ad}{ij}g_{bdck} 
                      - \sum_l \ta{ab}{il}g_{ljck}\Big)$

            \State $\tta{abc}{ijk} \gets 4\ta{abc}{ijk}
                                       - 2\ta{acb}{ijk}
                                       - 2\ta{cba}{ijk}
                                       - 2\ta{bac}{ijk}
                                       +  \ta{bca}{ijk}
                                       +  \ta{cab}{ijk}$

            \For{Permutations of $i,j,k$}
               \State $\tZv{abid} \meq \sum_c \tta{abc}{ijk}g_{jdkc}$
               \State $\tZo{ajil} \peq \sum_{bc} \tta{abc}{ijk}g_{lbkc}$
            \EndFor

         \EndFor
      \EndFor
   \EndFor
\end{algorithmic}
\end{algorithm}

The algorithm for the Jacobian transformation of a trial vector,
see supporting information,  
resembles the algorithm for the ground state,
but it is separated into two loops.
In the first, 
$\bt_{3}$ is constructed and contracted with an $\bR_1$-dependent intermediate.
In the second loop,
the routine used to construct $\bt_{3}$ is used again, 
but called twice with different input tensors to construct $\bR_3$.
The excitation vector is then transformed to contravariant form and 
contracted with the same integrals as the ground state 
to construct the excited state residual vector.

The algorithm for the transpose Jacobian transformation,  
is similar to the right transformation.
First, 
the $\bt_{3}$ amplitudes are computed and contracted in a separate loop over $i,j,k$,
before the main loop,
where the contribution of the $\bL_3$ amplitudes are calculated.
The contributions to the left Jacobian transform should be constructed from the 
contravariant form of $\bL_3$.
However, constructing the contravariant form directly
is complicated and requires several expensive linear combinations. 
The covariant form, 
on the other hand, 
can be constructed using contractions similar to those required for $\bt_3$
and six outer products,
avoiding any linear combinations. 
The contravariant form is then obtained using eq\,\eqref{contra_t}.
A complication for the transpose transformation 
is that it requires the construction of intermediates inside the $i,j,k$ loop.
One of these intermediates requires $\nv{3}\no{}$ memory 
and we have to add batching functionality,
writing and reading the intermediate from file for each batch.
To avoid construction of the full $\nv{4}$ integrals, 
the intermediates are contracted directly with Cholesky vectors outside the $i,j,k$ loop. 
Asymptotically, 
the computational cost is $4\nv{4}\no{3}$ FLOP,
the same as for the right transformation.

\begin{algorithm}
\caption{Algorithm to compute the CC3 contribution to $\BD^{m,0}$.}
\label{alg_ltdm}
\begin{algorithmic}\onehalfspacing
   \For{i = 1, $\no{}$}
      \For{j = 1, i}
         \For{k = 1, j}

            \State $\ta{abc}{ijk} \gets - (\eps{abc}{ijk})^{-1} P^{abc}_{ijk}
                   \Big(\sum_d \ta{ad}{ij}g_{bdck} - \sum_l \ta{ab}{il}g_{ljck}\Big)$

            \State $\Lx{abc}{ijk} \gets (\omega - \eps{abc}{ijk})^{-1} P^{abc}_{ijk}\Big(
                                    \Lx{a}{i}g_{jbkc} + \Lx{ab}{ij}F_{kc}
                           + \sum_d \Lx{ad}{jk}g_{ibdc} - \sum_l \Lx{ab}{lk}g_{iljc}\Big)$

            \State $\tLx{abc}{ijk} \gets 4\Lx{abc}{ijk}
                                       - 2\Lx{acb}{ijk}
                                       - 2\Lx{cba}{ijk}
                                       - 2\Lx{bac}{ijk}
                                       +  \Lx{bca}{ijk}
                                       +  \Lx{cab}{ijk}$

            \For{Permutations of $i,j,k$}
               \State $Y^{\text{o}}_{clik} \peq \sum_{ab} \tLx{abc}{ijk} \ta{ab}{lj}$
               \State $D^{m,0}_{cd} \peq \half\sum_{ab}\tLx{abc}{ijk}\ta{abd}{ijk}$
            \EndFor

            \State $\tta{abc}{ijk} \gets 4\ta{abc}{ijk}
                                       - 2\ta{acb}{ijk}
                                       - 2\ta{cba}{ijk}
                                       - 2\ta{bac}{ijk}
                                       +  \ta{bca}{ijk}
                                       +  \ta{cab}{ijk}$
            \For{Permutations of $i,j,k$}
               \State $D^{m,0}_{kc} \peq \sum_{ab}\Lx{ab}{ij}\tta{abc}{ijk}$
            \EndFor

         \EndFor
       \EndFor
   \EndFor
   \\
   \State $D^{m,0}_{ld} \meq \ssum{c}{ik} Y^\text{o}_{clik} \tau^{cd}_{ki}$
   \\
   \For{a = 1, $\nv{}$}
      \For{b = 1, a}
         \For{c = 1, b}

            \State $\tau^{abc}_{ijk} \gets - (\eps{abc}{ijk})^{-1} P^{abc}_{ijk}
                   \Big(\sum_d \tau^{ad}_{ij}g_{bdck} -
                        \sum_l \tau^{ab}_{il}g_{ljck}\Big)$

            \State $\Lx{abc}{ijk} \gets (\omega - \eps{abc}{ijk})^{-1} P^{abc}_{ijk}\Big(
                                    \Lx{a}{i}g_{jbkc} + \Lx{ab}{ij}F_{kc}
                           + \sum_d \Lx{ad}{jk}g_{ibdc} - \sum_l \Lx{ab}{lk}g_{iljc}\Big)$

            \State $\tLx{abc}{ijk} \gets 4\Lx{abc}{ijk}
                                       - 2\Lx{acb}{ijk}
                                       - 2\Lx{cba}{ijk}
                                       - 2\Lx{bac}{ijk}
                                       +  \Lx{bca}{ijk}
                                       +  \Lx{cab}{ijk}$

            \For{Permutations of $a,b,c$}
               \State $D^{m,0}_{lk} \meq \half\sum_{ij}\tLx{abc}{ijk}\ta{abc}{ijl}$
            \EndFor

         \EndFor
      \EndFor
   \EndFor
\end{algorithmic}
\end{algorithm}

In Algorithm\,\ref{alg_ltdm}, we show how to compute the $\bL_3$ 
contributions to $\BD^{m,0}$,
see eq\,\eqref{D}. 
The same algorithm can be used to compute the ground state density, 
$\BD^{0,0}$, by inserting $\bLa_3$ instead of $\bL_3$.
For $\tBD^{0,m}$ several intermediates from the ground state density, 
as well as the ground state density itself, 
are reused, see the supporting information.

The main difference between Algorithm\,\ref{alg_ltdm} 
and the algorithm for the Jacobian transformations
is the additional triple loop over the virtual indices.
This loop is required due to the occupied-occupied block of the density matrix 
that has contributions from two triples tensors with different occupied indices.
Therefore, 
it is not possible to use the previous scheme of holding only 
triples amplitudes for a given $i,j,k$.
In a CC3 calculation, 
the number of virtual orbitals is much larger than the number of occupied orbitals  
when a reasonable basis set is used.
Therefore, 
the BLAS\cite{BLAS, BLAS3} routines do not parallelize well,
and the serial loop over the virtual indices would be inefficient. 
To circumvent this, 
the loops over the virtual indices were parallelized
using OpenMP\cite{openmp18}.
The triples tensors have to be constructed once for
fixed occupied and once for fixed virtual indices and
the computational cost of constructing the CC3 transition densities
increases to 13$\nv{4}\no{3}$ per state.
Nevertheless, 
the construction of the densities constitutes
only a small fraction of the time compared to the iterative solution
of the excited state equations.

\section{Applications}\label{Applications}

To demonstrate the performance of the code,
we have calculated the two lowest CC3 singlet valence excited states of acetamide
using aug-cc-pVDZ\cite{aug_cc_pVXZ} with $e^T$, \psif\cite{Psi4}, \cfour,\cite{CFOUR,CFOURPara,CC3resp}
and the two implementations in \dalton\cite{Dalton,Dalton2018,CC3imp}.
The timing data and the number of iterations for converging the ground state
and both excited states are summarized in Table \ref{tab_comparison}.
\begin{table}
   \centering
   \caption{Timings to compare \cfour, \dalton{}, $e^T$ and \psif.
            The timings reported refer to one iteration of the ground state equations,
            one iteration of the excited state equations and the total execution time.
            For the ground state and the excited states $n_{\text{Iter}}$ 
            specifies the number of iterations to converge the respective states.
            The calculations were performed on one node with four Intel Xeon Gold 6130 CPU 
            with 16 cores each using 40 cores and using a total of 180 GB shared memory.}
   \label{tab_comparison}
   \begin{tabular}{l S[table-format=3] S[table-format=2] S[table-format=4] c S[table-format=4]}
   \toprule
	& 
   \mc{2}{c}{Ground state} & 
   \mc{2}{c}{Excited states} & 
   \mc{1}{c}{Total} \\ 
   \cmidrule(lr){2-3} \cmidrule(lr){4-5} \cmidrule(lr){6-6}
   & \mc{1}{c}{wall time [s]} & \mc{1}{c}{$n_{\text{Iter}}$} 
   & \mc{1}{c}{wall time [s]} & \mc{1}{c}{$n_{\text{Iter}}$} 
   & \mc{1}{c}{wall time [min]} \\
   \midrule
   $e^T$	               &  16 & 13 &   28 & 65 &   34 \\
   \dalton{} new        &  47 & 13 &   97 & 62 &  129 \\
   \cfour{} symmetry    & 150 & 13 &  320 & 38 &  240 \\
   \cfour{} no symmetry & 330 & 13 &  685 & 34 &  468 \\
   \dalton{} old	      & 267 & 13 &  767 & 71 &  971 \\
   \psif	               & 404 &  8 & 1187 & 49 & 1040 \\
   \bottomrule
\end{tabular}
\end{table}
When running \cfour{}, 
the oldest \dalton{} implementation and \psif{} the $C_S$ symmetry of acetamide has been exploited.
For comparison \cfour{} was also run without symmetry.
The threshold for the convergence of the ground state residual was $10^{-6}$,
while we used a threshold of $10^{-4}$ for the excited states.
With \psif{}, 
both the ground and excited state residuals were converged to $10^{-4}$, 
which is why only eight iterations were needed to converge $\bOm$.
The differences in the convergence of the excited state equations are due to
the different start guesses the programs use.
While \psif{} first converges the CCSD equations and restarts CC3 from CCSD,
the other programs use orbital energy differences as default start guesses.
Note that all these programs can restart from the CCSD solution, 
but it is the default behavior of \psif.
As the lowest excited state is not dominated by the lowest orbital energy difference
a specific start guess had to be chosen to obtain the lowest root with \cfour{}.
This start guess improved the convergence behavior of \cfour{} significantly.
To remove the dependence on the number of iterations,
we report timings per iteration which are dominated by the time spent
computing the CC3 contributions.
However, 
\psif{} does not report timings per iteration to converge the ground state equations 
and \cfour{} does not report timings per iteration for converging the excited state
equations.
Therefore,
the total time spent solving for the ground state and the excited states,
respectively,
was divided by the number of iterations.
Even though the reported timings might not compare entirely identical steps in the codes,
Table \ref{tab_comparison} clearly shows the efficiency of the CC3 code in $e^T$. 

To demonstrate the capabilities of the code, 
we have calculated singlet valence and core excitation energies
and EOM oscillator strengths for the amino acid L-proline
(C$_5$H$_9$NO$_2$)\cite{pubchemproline}.
One valence excitation energy was calculated at the CCSD/aug-cc-pVTZ 
and CC3/aug-cc-pVTZ levels of theory using the frozen core (FC) approximation
resulting in 23 occupied and 544 virtual orbitals.\cite{aug_cc_pVXZ}

Table\,\ref{tab_valence} shows the excitation energy and oscillator strength
for the lowest valence excited state at the CCSD and CC3 level.
The excitation vector has 96\% singles contribution and the
excitation energies differ by about 0.11\,eV.
In Table\,\ref{tab_valence_time},
we report the averaged time per routine call 
as well as an estimate for the computational efficiency 
and the number of routine calls.
For the ground state,
for example,
$\ncall$ specifies the number of times 
the ground state residual vector is computed.
The efficiency is defined as the observed FLOP per second (FLOPS) divided 
by the theoretical maximum number of FLOPS. 
For the CPUs used for this calculation, 
two Intel Xeon Gold 6152 processors, 
the theoretical maximum is given by\cite{inteldata} 
\begin{equation}
    2 \mathrm{CPUs} \times 
    22 \mathrm{cores/CPU} \times 
    2.1 \mathrm{GHz} \times
    32 \mathrm{FLOP/cycle}
    = 2956.8 \mathrm{GFLOPS}
\end{equation}
When calculating the number of FLOP, 
we only count the dominant matrix-matrix multiplications 
with a FLOP cost of $2\nv{4}\no{3}$.
This will be an undercount of the total FLOP, 
but should give a ballpark estimate. 
Note that the CPUs have turbo boost technology, 
giving a maximum theoretical frequency of 3.7 GHz when one core is active 
and 2.8 GHz when 22 cores are active. 
For the highly efficient BLAS routines used for the matrix multiplications, 
the actual frequency is likely to be close to the base frequency of 2.1 GHz, however.

\begin{table}
   \centering
   \caption{Proline excitation energy and oscillator strength for the lowest
   singlet valence excitation at the CCSD and CC3 levels of theory.}
   \label{tab_valence}
   \begin{tabular}{c
                   S[round-mode=places, round-precision=2, table-format=1.2, table-number-alignment=center]
                   c 
                   c
                   S[round-mode=places, round-precision=2, table-format=1.2, table-number-alignment=center]}
   \toprule
   \multicolumn{2}{c}{CCSD} && \multicolumn{2}{c}{CC3}    \\
   $\omega$ [eV] & ${f\times100}$ && $\omega$ [eV] & ${f\times100}$ \\
   \midrule
   5.830      &  0.0775   &&   5.718    &    0.0661 \\
   \bottomrule
   \end{tabular}
\end{table}

\begin{table}
   \centering
   \caption{Timings for the different parts of the calculation of one valence excited state
            with oscillator strengths in L-proline at the CC3 level of theory.
            $\ncall$ specifies the number of calls to the subroutines constructing
            the respective quantity.
            Timings have been averaged over the number of routine calls.
            The calculations were performed on one node
            with two Intel Xeon Gold 6152 processors with 22 cores each
            and using a total of 700 GB shared memory.}
   \label{tab_valence_time}
   \begin{tabular}{lccS[table-format=3]}
   \toprule
   Contributions   & 
   wall time [min]  & 
   efficiency [\%] & 
   \mc{1}{c}{$\ncall$} \\
   \midrule
   Ground state               & 163 & 14.7 & 10 \\
   Prepare for multipliers    & 169 & 14.2 &  1 \\
   Multipliers                & 347 & 13.8 & 11 \\
   Prepare for Jacobian       & 147 & 16.3 &  1 \\
   Right excited states       & 281 & 17.1 & 26 \\
   Prepare for Jacobian       & 160 & 15.0 &  1 \\
   Left excited states        & 341 & 14.1 & 28 \\
   $\BD^{0,0}$                & 379 & 15.8 &  1 \\
   $\BD^{m,0}$                & 382 & 15.7 &  1 \\
   $\tBD^{0,m}$               & 530 & 15.9 &  1 \\
   \bottomrule
\end{tabular}
\end{table}

From Table\,\ref{tab_valence_time} we observe that one iteration of
the multiplier equations is approximately twice as expensive as one iteration
for the ground state.
The transpose Jacobian transformation,
which is required for the multipliers,
costs 8$\nv{4}\no{3}$ FLOP compared to 4$\nv{4}\no{3}$ FLOP
for the ground state.
The timings to obtain left excited states are roughly the same as the timings
to solve for the multipliers because a trial vector is transformed
by the transpose of the Jacobian.
Note that the timings in Table\,\ref{tab_valence_time} were obtained 
with an older version of the code 
that required the construction of the full $\nv{4}$ integrals for the left vectors 
and did not exploit the covariant-contravariant transformations. 
In the preparation routines the intermediates used in the Jacobian transformations
are computed, as shown in Algorithm \ref{alg_prep}.
The preparation is as expensive as one iteration for the ground state,
but we save $2\nv{4}\no{3}$ FLOP per Jacobian transformation.
The ground state density and $\BD^{m,0}$ are calculated using the same routines
and the computational cost is the same.
The CC3 contribution to $\tBD^{0,m}$ requires $\bt_{3}$, $\bla_{3}$ and $\bR_3$.
In addition, 
$\bR_3$ is approximately twice as expensive to compute as $\bt_{3}$,
so $\tBD^{0,m}$ is considerably more expensive than $\BD^{m,0}$.

\begin{figure}[ht]
   \centering
   \includegraphics[width=\textwidth]{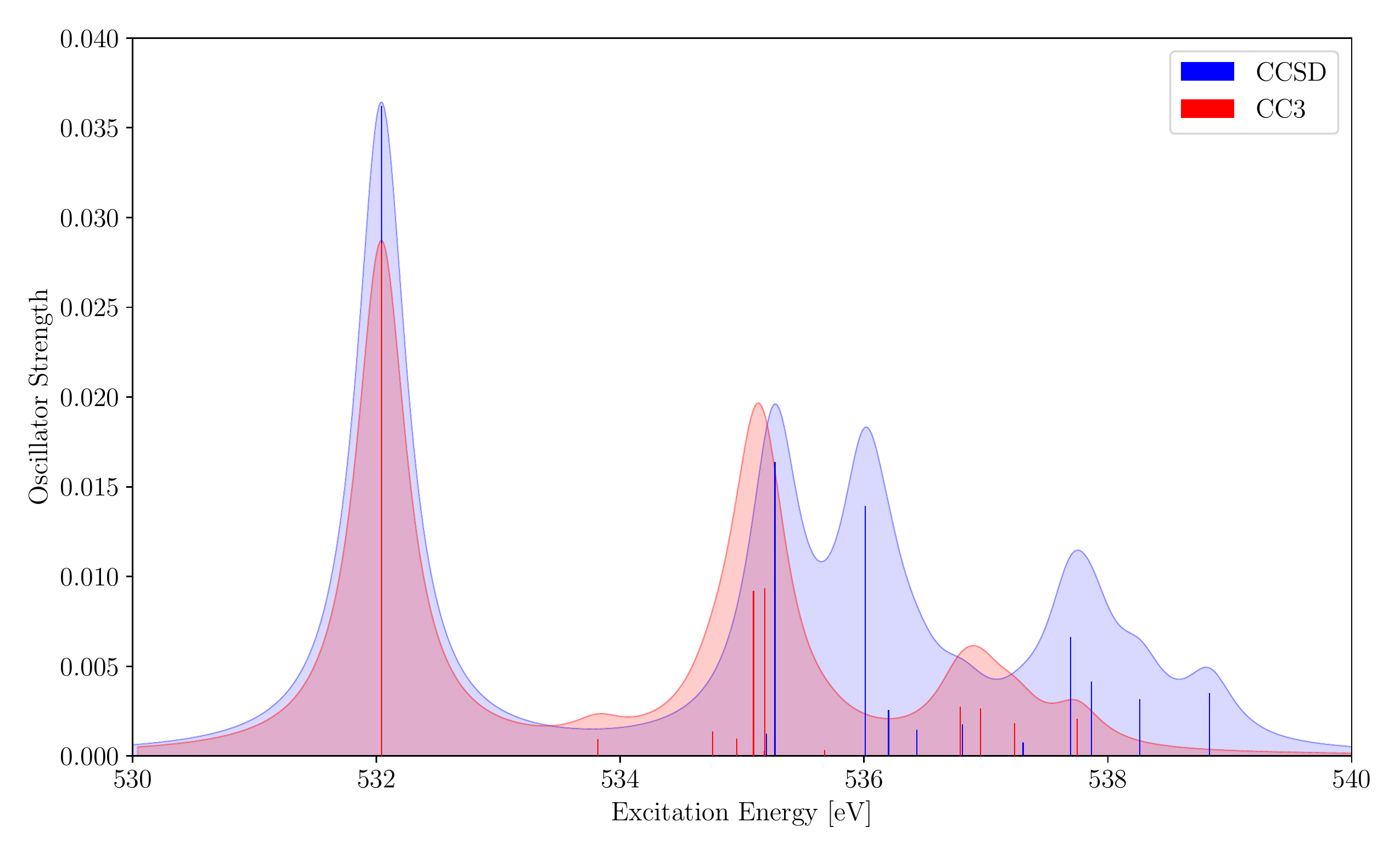}
   \caption{Core excitation spectrum of the oxygen atoms 
   of L-proline computed with CC3 (red) and CCSD (blue).
   The peaks were broadened using a Lorentzian line shape and a width of 0.5 eV. 
   The CCSD spectrum is shifted by -1.9\,eV to match 
   the first peak of the CC3 spectrum.}
   \label{spectrum}
\end{figure}

We have also calculated six core excited states for each of the oxygen atoms,
using core valence separation (CVS).
The aug-cc-pCVTZ basis set was used on the oxygen atom that was excited and
aug-cc-pVDZ for the rest of the molecule
(31 occupied and 270 virtual orbitals).\cite{aug_cc_pVXZ,cc_pCVXZ}
In Table\,\ref{tab_CO-edge} we show the results for core excitations
from the carbonyl oxygen of L-proline.
Due to the better description of relaxation effects by the inclusion of triple
excitations, 
the excitation energies obtained with CC3 are up to 3 eV
lower than the corresponding CCSD excitation energies.
The same trends are observed for core excitations from the hydroxyl oxygen,
as shown in Table\,\ref{tab_OH-edge}.
The CC3 oscillator strengths are between 16\% and 60\% lower than the
values obtained with CCSD.
In Figure\,\ref{spectrum} we show NEXAFS spectra computed with EOM-CCSD and EOM-CC3.
Despite shifting the CCSD spectrum by $-1.9$\,eV,
the two spectra show significant differences.
From the CCSD spectrum one would expect two peaks between 535\,eV and 536\,eV
and the peak at 534\,eV is not present in the shifted CCSD plot. 
The calculated CC3 excitation energies are in good agreement with experimental data
reported by Plekan \textit{et al.} in Ref.\,\citenum{PlekanGly1}.
The authors measured the first excitation from the carbonyl oxygen at 532.2\,eV
and a broad peak from the hydroxyl oxygen at 535.4\,eV,
consistent with the first two calculated CC3 excitation energies.
Note that taking relativistic effects into account 
will increase the excitation energies by about 0.38\,eV,
while increasing the basis set would lower them somewhat\cite{cvs_basis,glycine}.

\begin{table}
   \centering
   \caption{Proline excitation energies and oscillator strengths 
   for core excitations from the carbonyl oxygen at the CCSD and CC3 level of theory.}
   \label{tab_CO-edge}
   \begin{tabular}{c
                   S[round-mode=places, round-precision=2, table-format=1.2, table-number-alignment=center]
                   c 
                   c
                   S[round-mode=places, round-precision=2, table-format=1.2, table-number-alignment=center]}
   \toprule
   \multicolumn{2}{c}{CCSD}  && \multicolumn{2}{c}{CC3}   \\
   $\omega$ [eV] & ${f\times100}$ && $\omega$ [eV] & ${f\times100}$ \\
   \midrule
   533.943 & 3.6218 && 532.040 & 2.8539 \\
   537.103 & 0.1238 && 533.817 & 0.0953 \\
   538.104 & 0.2566 && 534.756 & 0.1377 \\
   538.335 & 0.1477 && 534.953 & 0.0986 \\
   538.710 & 0.1779 && 535.179 & 0.0262 \\
   539.207 & 0.0761 && 535.677 & 0.0340 \\
   \bottomrule
   \end{tabular}
\end{table}

\begin{table}
   \centering
   \caption{Proline excitation energies and oscillator strengths 
   for core excitations from the hydroxyl oxygen at the CCSD and CC3 level of theory.}
   \label{tab_OH-edge}
   \begin{tabular}{c
                   S[round-mode=places, round-precision=2, table-format=1.2, table-number-alignment=center]
                   c 
                   c
                   S[round-mode=places, round-precision=2, table-format=1.2, table-number-alignment=center]}
   \toprule
   \multicolumn{2}{c}{CCSD}  && \multicolumn{2}{c}{CC3}   \\
   $\omega$ [eV] & ${f\times100}$ && $\omega$ [eV] & ${f\times100}$ \\
   \midrule
   537.172 & 1.6373 && 535.093 & 0.9192 \\
   537.911 & 1.3923 && 535.186 & 0.9351 \\
   539.598 & 0.6640 && 536.789 & 0.2758 \\
   539.770 & 0.4145 && 536.955 & 0.2643 \\
   540.165 & 0.3164 && 537.235 & 0.1824 \\
   540.736 & 0.3508 && 537.747 & 0.2088 \\
   \bottomrule
   \end{tabular}
\end{table}

Timings for the calculations of the core excited states are reported in
Table\,\ref{tab_CO-edge_time} for excitations from the carbonyl oxygen.
The timings for the core excitations from the hydroxyl oxygen
are not reported because they are almost identical.
Compared to the valence excited state calculation,  
the timings for the ground state and the
multipliers are reduced due to the use of smaller basis sets.
The CVS approximation reduces the computational cost 
of the Jacobian transformations
from 8$\nv{4}\no{3}$ to 8$\nv{4}\no{2}$ FLOP\cite{thy_dyn, CCaccuracy5}.
Therefore, one iteration is 6 times faster than a ground state iteration. 
These savings are achieved by cycling the triple loop over the occupied indices
when none of the indices correspond to the core orbitals of interest.
Similar savings can be achieved during the construction of the transition densities.
However,
in the present implementation only the triple loop
over the occupied indices can be cycled
but not the loops over the virtual indices.
The efficiency is improved compared to the valence excitation calculation
as the contravariant code was used for this calculation.

\begin{table}
   \centering
   \caption{Timings for the different parts of the calculation of 6 core excited states
            (located at the carbonyl oxygen) with oscillator strengths for L-proline
            at the CC3 level of theory. $\ncall$ specifies the number of calls
            to the subroutines constructing the respective quantities.
            Timings have been averaged over the number of routine calls.
            The calculations were performed on nodes 
            with two Intel Xeon Gold 6138 processors with 20 cores each
            and using a total of 370 GB shared memory.}
   \label{tab_CO-edge_time}
   \begin{tabular}{lS[table-format=3]S[table-format=3.1]S[table-format=4]}
   \toprule
   Contributions                 & 
   \mc{1}{c}{wall time [min]}    & 
   \mc{1}{c}{efficiency [\%]}    & 
   \mc{1}{c}{$\ncall$} \\
   \midrule
   Ground state            & 19 & 22.2 &  12 \\
   Prepare for multipliers & 17 & 24.8 &   1 \\
   Multipliers             & 31 & 26.4 &  14 \\
   Prepare for Jacobian    & 16 & 25.2 &   1 \\
   Right excited states    &  3 &  7.7 & 290 \\
   Left excited states     &  3 &  8.4 & 315 \\
   $\BD^{0,0}$             & 50 & 10.2 &   1 \\
   $\BD^{m,0}$             & 23 &  8.8 &   6 \\
   $\tBD^{0,m}$            & 45 &  6.9 &   6 \\
   \bottomrule
\end{tabular}
\end{table}

In Table\,\ref{parallel}, 
we present timings from calculations on furan with aug-cc-pVDZ basis set,
using 1, 5, 10, 20 and 40 threads.
We calculated the transition moments 
from the ground state 
to the first excited state, 
which requires solving for $\bs{\tau}$, $\bs{\lambda}$, $\bR$ and $\bL$.
We also report speedups relative to the single thread calculation.
Increasing the number of threads from 1 to 40 reduces 
the total wall time by approximately a factor of 15.
Due to dynamic overclocking, 
the theoretical maximum frequency for the single threaded case is 3.7 GHz
while it is 2.7 GHz with 20 active cores per processor 
and the base frequency is 2.0 GHz\cite{inteldata}.

\begin{table}
   \centering
   \caption{Timings for calculating the EOM transition moment 
            for the first excited state of furan in seconds
            using 1, 5, 10, 20 and 40 threads. 
            Total times as well as timings for solving for
            $\bs{\tau}$, $\bs{\lambda}$, $\bR$ and $\bL$
            are reported.
            Numbers of iterations are given in parentheses.
            The speedup compared to a single core is given next to the timing.
            The remaining time is primarily spent constructing 
            the density matrices.
            The calculations were performed on a node 
            with two Intel Xeon Gold 6138 2.0 GHz processors with 20 cores each
            and using a total of 150 GB shared memory.}
   \label{parallel}
   \begin{tabular}{S[table-format=3]
                   S[table-format=5]S[table-format=2.2]
                   S[table-format=4]S[table-format=2.2]
                   S[table-format=4]S[table-format=2.2]
                   S[table-format=4]S[table-format=2.2]
                   S[table-format=4]S[table-format=2.2]}
      \toprule
      \mc{1}{c}{Threads}             &
      \mc{2}{c}{total}               &
      \mc{2}{c}{$\bs{\tau}$ (13)}    &
      \mc{2}{c}{$\bs{\lambda}$ (14)} &
      \mc{2}{c}{$\bR$ (15)}          &
      \mc{2}{c}{$\bL$ (16)}          \\
      \midrule
       1 & 35197 &  \nan & 4231 &  \nan & 8403 &  \nan & 9148 &  \nan & 9605 &  \nan \\
       5 &  8630 &  4.08 & 1067 &  3.96 & 2059 &  4.08 & 2259 &  4.05 & 2347 &  4.09 \\
      10 &  4612 &  7.63 &  572 &  7.40 & 1103 &  7.62 & 1199 &  7.63 & 1252 &  7.67 \\
      20 &  2841 & 12.39 &  353 & 11.98 &  691 & 12.16 &  743 & 12.32 &  763 & 12.58 \\
      40 &  2286 & 15.39 &  290 & 14.61 &  563 & 14.94 &  587 & 15.60 &  632 & 15.19 \\
      \bottomrule
   \end{tabular}
\end{table}

\begin{figure}[ht]
   \centering
   \begin{subfigure}{0.4\textwidth}
      \includegraphics[width=\textwidth]{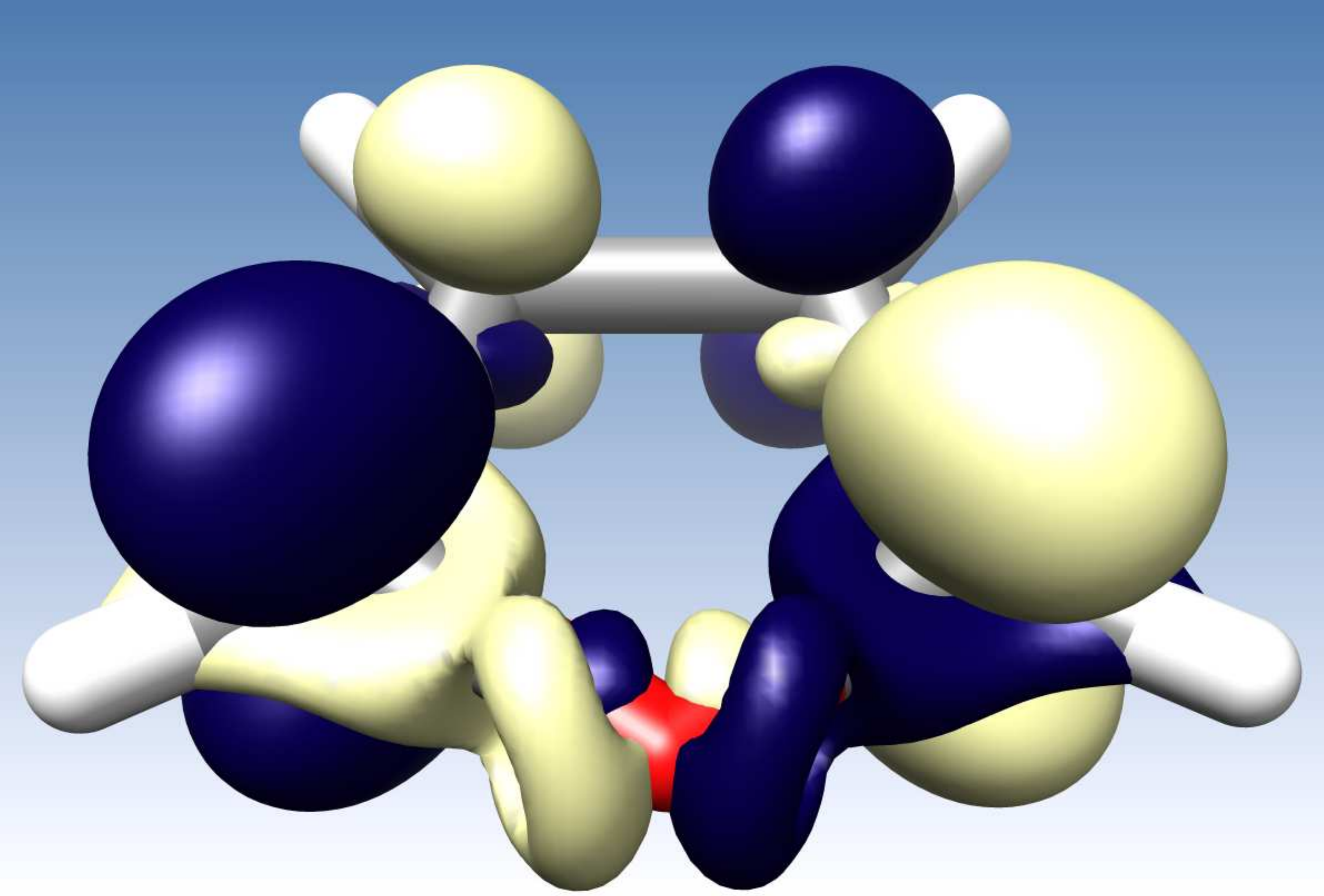}
      \caption{$\tBD^{0,2}_{CC3}$}
      \label{td}
   \end{subfigure}
   \begin{subfigure}{0.4\textwidth}
      \includegraphics[width=\textwidth]{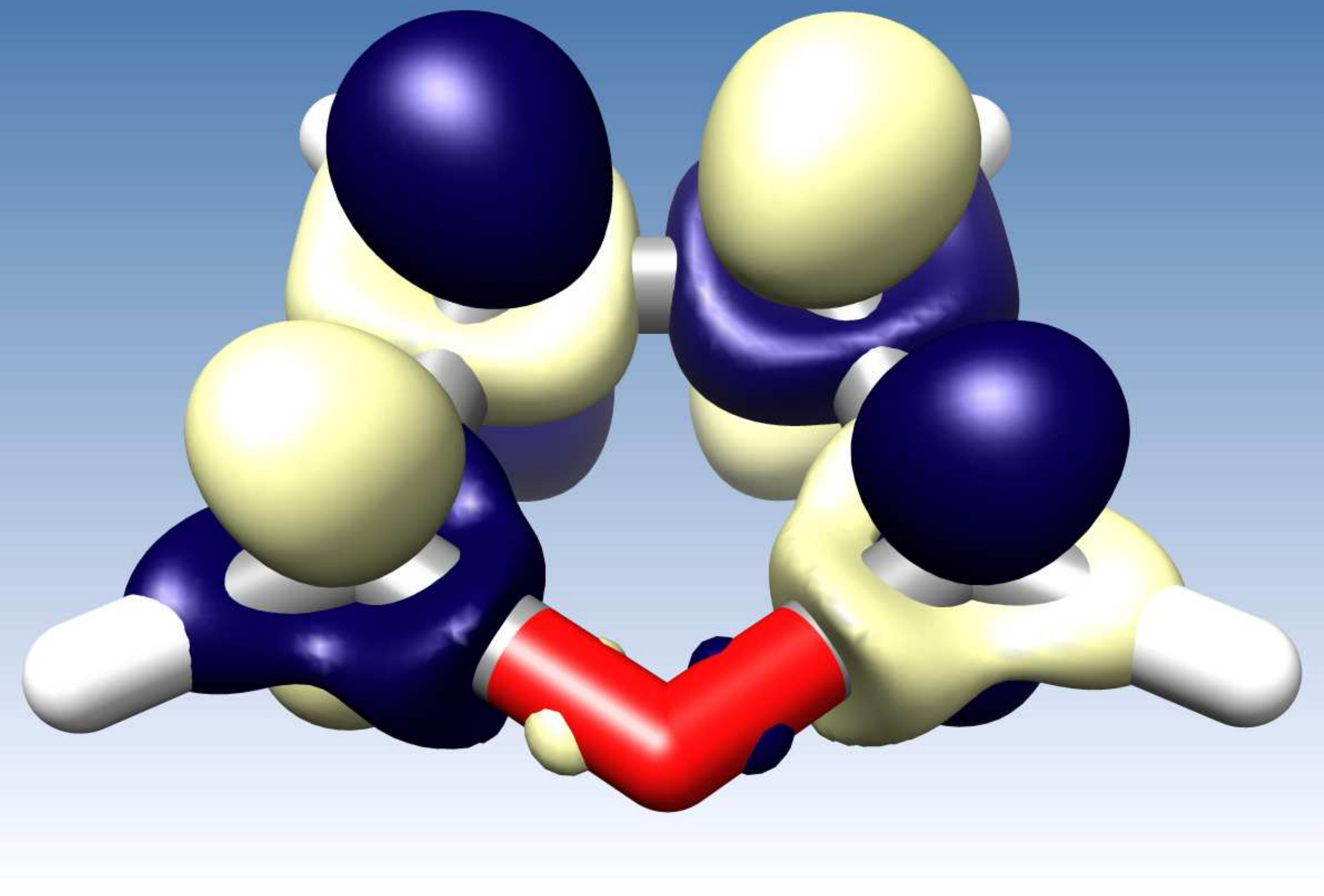}
      \caption{$\tBD^{0,2}_{CC3} - \tBD^{0,2}_{CCSD}$}
      \label{tddiff}
   \end{subfigure}
   \caption{Transition densities of furan.
            \subref{td} 
            The second CC3 transition density 
            ($\tBD{}{}^{CC3}$)
            with contour value 0.006.
            \subref{tddiff}
            Difference from CCSD 
            ($\tBD^{0,2}_{CC3} - \tBD^{0,2}_{CCSD}$)
            with contour value 0.0003.}
   \label{transdens}
\end{figure}

Finally, in Figure\,\ref{transdens}, 
we show the CC3 transition density, $\tBD^{0,2}_{CC3}$,
as well as the difference between the CC3 and CCSD transition densities,
$\tBD^{0,2}_{CC3} - \tBD^{0,2}_{CCSD}$,
plotted using Chimera\cite{Chimera}. 
While the difference between the densities is small
(the contour value is only 0.0003),
the triples decrease the volume at the same contour value of the transition density,
which goes along with an increase in the double excitation character.
This is reflected in a reduction of the oscillator strength from 0.181 to 0.168
and an increase in the weight of the doubles in the excitation vectors.

\section{Conclusion}\label{conclusion}

In this paper we have described an efficient implementation of the CC3 model
including ground state and excited state energies as well as EOM oscillator strengths.
To the best of our knowledge, the algorithm reported is the most efficient for
canonical CC3 and the first implementation of EOM-CC3 transition densities.
The computational cost of excited states is reduced to 8$\nv{4}\no{3}$ FLOP,
due to the introduction of intermediates constructed outside the iterative loop.
The code is parallelized using OpenMP 
and the algorithm can be extended to utilize MPI through coarrays
which are included in the Fortran 2008 standard.

A possible modification of the code is to use triple loops over the virtual orbitals 
for the construction of the amplitudes. 
OpenMP parallelization will then happen at the level of the triple loops, 
which is already implemented for parts of the density construction. 
Early experimental code indicates that the efficiency of the matrix-matrix 
multiplications are then somewhat reduced, 
but the overhead due to reordering almost vanishes. 
This is probably related to the spatial locality of the arrays in memory.
Another advantage of such a scheme is that it can be adapted for graphical processing units. 

Finally, 
the extension to the densities of excited states 
and the transition densities between excited states
is straightforward and will be reported elsewhere. 


\begin{acknowledgement}

We thank reviewer 1 for suggesting the use of the contravariant-covariant transformation. 
We acknowledge computing resources through UNINETT Sigma2
- the National Infrastructure for High Performance Computing
and Data Storage in Norway, through project number NN2962k.
We acknowledge funding from the Marie Sk\l odowska-Curie European Training Network
``COSINE -- COmputational Spectroscopy In Natural sciences and Engineering'',
Grant Agreement No. 765739 and
the Research Council of Norway through FRINATEK projects 263110, CCGPU,
and 275506, TheoLight.

\end{acknowledgement}


\clearpage


\bibliography{cc3}

\end{document}